# 3D Brainformer: 3D Fusion Transformer for Brain Tumor Segmentation

Rui Nian, Guoyao Zhang, Yao Sui, Yuqi Qian, Qiuying Li, Mingzhang Zhao, Jianhui Li, Ali Gholipour, and Simon K. Warfield

*Abstract*—**Magnetic resonance imaging (MRI) is critically important for brain mapping in both scientific research and clinical studies. Precise segmentation of brain tumors facilitates clinical diagnosis, evaluations, and surgical planning. Deep learning has recently emerged to improve brain tumor segmentation and achieved impressive results. Convolutional architectures are widely used to implement those neural networks. By the nature of limited receptive fields, however, those architectures are subject to representing long-range spatial dependencies of the voxel intensities in MRI images. Transformers have been leveraged recently to address the above limitations of convolutional networks. Unfortunately, the majority of current Transformers-based methods in segmentation are performed with 2D MRI slices, instead of 3D volumes. Moreover, it is difficult to incorporate the structures between layers because each head is calculated independently in the Multi-Head Self-Attention mechanism (MHSA). In this work, we proposed a 3D Transformer-based segmentation approach. We developed a Fusion-Head Self-Attention mechanism (FHSA) to combine each attention head through attention logic and weight mapping, for the exploration of the long-range spatial dependencies in 3D MRI images. We implemented a plug-and-play self-attention module, named the Infinite Deformable Fusion Transformer Module (IDFTM), to extract features on any deformable feature maps. We applied our approach to the task of brain tumor segmentation, and assessed it on the public BRATS datasets. The experimental results demonstrated that our proposed approach achieved superior performance, in comparison to several state-of-the-art segmentation methods.**

*Index Terms*—**Magnetic Resonance Imaging, Transformer, Self-Attention, Convolution, Brain Tumor Segmentation.**

Rui Nian, Guoyao Zhang, Yuqi Qian, Qiuying Li, Mingzhang Zhao, Jianhui Li are with the School of Electronic Engineering, Ocean University of China, Qingdao, China. (Corresponding author: Rui Nian. E-mail: nianrui_80@163.com)

Yao Sui is with National Institute of Health Data Science, Peking University, Beijing, China.

Rui Nian, Ali Gholipour, Simon K. Warfield are with Harvard Medical School, and with Computational Radiology Laboratory, Boston Children's Hospital, Boston, MA, United States.

## I. INTRODUCTION

MAGNETIC Resonance Imaging (MRI) plays a critically important role in scientific research and clinical studies. It is more prominent in brain mapping for tasks, such as delineating anatomies, explaining cognition schemes, and diagnosing brain diseases. As a safe, non-invasive, accurate, and in-vivo imaging technique, MRI is typically applied in the diagnosis of brain tumors for clinical decisions.

### A. Background and Significance

MRI imaging allows for detecting brain tumors reliably, measuring brain tumors accurately, and localizing brain tumors precisely, and in turn, enables effective pre-surgical plans and post-surgical outcome evaluations. In clinical routines, brain tumor are manually segmented into heterogeneous sub-regions, i.e., edema (ED), enhancing tumor (ET), non-enhancing tumor core (NET) and necrotic (NCR), when referring to MRI acquisitions with different contrast levels. Since MRI imaging is 3D volumetric and comprises a large number of voxels, manual segmentation is time-consuming, costly, less efficient, and requires clinical expertise. Clinicians analyze MRI imaging often by 2D slices in planes so that a reduced number of voxels could be handled. However, such a strategy loses too much 3D volumetric information, leading to biased analysis. Consequently, 3D automatic brain tumor segmentation has emerged in recent decades, which yields accurate segmentation, while in parallel, enables efficient and reproducible results. The analysis over 3D volumes brings more insights to clinical decisions, as compared to those 2D slice-based methods [1]. It releases surgeons from the complicated pipelines for manual analysis, leading to reduced time and efforts in pre-surgical planning. Though it is desirable in clinical routines, it is still challenging to segment 3D brain tumor out from the normal anatomies, because (1) it is difficult to exploit prior information about the structure and locations of brain tumors in comparison to those widely-used robust segmentation algorithms of other anatomical structures, considering that the size, shape, and localization of brain tumors have considerable variations across patients; and (2) the boundaries between adjacent structures are often ambiguous due to the smooth intensity gradients, partial volume effects and bias field artifacts.

## B. State-of-The-Art in MRI Image Segmentation

A variety of machine learning-based methods have been developed to segment the abnormal brain tissues out from normal tissues with MRI images, including classification-based methods with support vector machines and decision tree. Deep learning based methods, implemented with convolutional neural networks (CNNs) [2], have recently emerged as a powerful tool for various medical image segmentation tasks. One of the most popular work that has been recognized recently is U-Net [3]. Many variants have subsequently developed based on U-Net architecture, including U-Net++ [4], UNet3+ [5], 3D U-Net [6], V-Net [7], Res-UNet [8], and Dense-UNet [9]. Although CNNs have made great success in the field of medical image computing, it is still difficult to make further breakthroughs, for example, since the intrinsic locality of convolution operation, it is difficult for CNN-based approaches to learn explicit global and long-range semantic information interaction. In addition, the size and shape of convolution kernels are typically fixed, so they cannot adapt to arbitrary input images [2]. Therefore, the segmentation of structures with various shapes and scales for brain tumors, is still suboptimal due to the nature of incapability in capturing multi-scale contextual information with CNNs. Studies have addressed this problem by using atrous convolutional layers, and pyramids. Unfortunately, these methods still have limitations in modeling spatially long-range dependencies.

The latest success in Transformers provides the design to exploit spatially long-range dependencies. The structure of Transformers is a Seq2Seq model consisting of encoder and decoder. Many variants have been recognized due to the big success of Transformers, such as Vision Transformer (ViT), DeiT, and DETR. Unfortunately, Transformers-based models have not yet attracted sufficient attention in medical image segmentation. Although a number of methods have been shown promising segmentation results, such as UNETR [10], TransUNet [11], MISSU [12], TransFuse and MedT, it is still challenging to segment brain tumor since most mis-segmented areas are located around the boundary of the region-of-interest. TransUNet has demonstrated that Transformers could be used as powerful encoders for medical image segmentation. TransFuse has been proposed to improve the efficiency for global context modeling by fusing Transformers with CNNs. MedT leveraged a Transformer axial attention mechanism for 2D medical image segmentation. However, these 2D networks are difficult to effectively capture the inter-layer information of brain tumors. UNETR has been built with a 3D network architecture, and used a pure Transformer as an encoder to learn the representation of the input 3D volumes. Unfortunately, it only applied MHSA mechanism to capture global multi-scale information in encoder, and kept the architecture with traditional CNNs in decoder. Therefore, it is difficult to establish complete multi-scale dependencies. MISSU proposed a new 3D medical image segmentation framework by Transformer, which connects a local multi-scale fusion block in the encoder and combines U-Net with Transformer to improve global contextual interactions and edge-detail preservation. However, this architecture still could not perform self-attention mechanism in the encoder-decoder and therefore could not fully establish remote spatial dependencies among voxels in medical volume.

## C. Contributions

There is an unmet need of technologies for accurate brain tumor segmentation with MRI images. In this paper, we elaborated in one deep learning based framework named 3D Brainformer that incorporates the advantages of the recently developed Transformers with the classic convolutional networks for 3D brain tumor segmentation tasks from volumetric MRI imaging. Specifically, we built an encoder with Fusion Transformer layers, which establishes the global connection among 3D volume data with Fusion-Head Self-Attention mechanism (FHSA); and then the design of Infinite Deformable Fusion Transformers Module (IDFTM) has been particularly put forward to act as a plug-and-play module that could be inserted into any position of the entire network architecture. We tried to break the bottlenecks to combine self-attention mechanism of long-range dependencies, together with the detailed adjacent correlation via Residual Basic Convolution Module (RBM). The thermal maps of the feature extraction process with the cascades of only RBMs, only IDFTMs, both IDFTM and RBM, are shown in Fig. 1 respectively. In our context, it could be observed that RBM paid more attention to the local characteristics of tumor core (TC) region in detail, while IDFTM focused more globally on ED region and ET region. We carried out extensive experiments to assess our approach. The experimental results demonstrated that our approach outperformed several state-of-the-art methods on the Multimodal Brain Tumor Image Segmentation Challenge (BraTS) datasets [1].

This work has three main contributions:
- We developed an self-attention mechanism that directly parse deeply encoded information in 3D MRI imaging, which is not subject to the sequentialized input in Transformers, and is able to perform feature extraction on arbitrarily variable feature maps.
- We proposed a hybrid Transformer strategy that applied self-attention mechanism by integrating multi-heads in 3D volumetric data at multi-scales in encoder.
- We leveraged a with 3D RBM and IDFTM in decoder, where RBM paid attention to local abstract details, and IDFTM is able to establish effective global dependencies from coarse and fine-grained feature representations with self-attention mechanism. Therefore, the fuzzy boundary of the brain tumor could be more accurately segmented out.

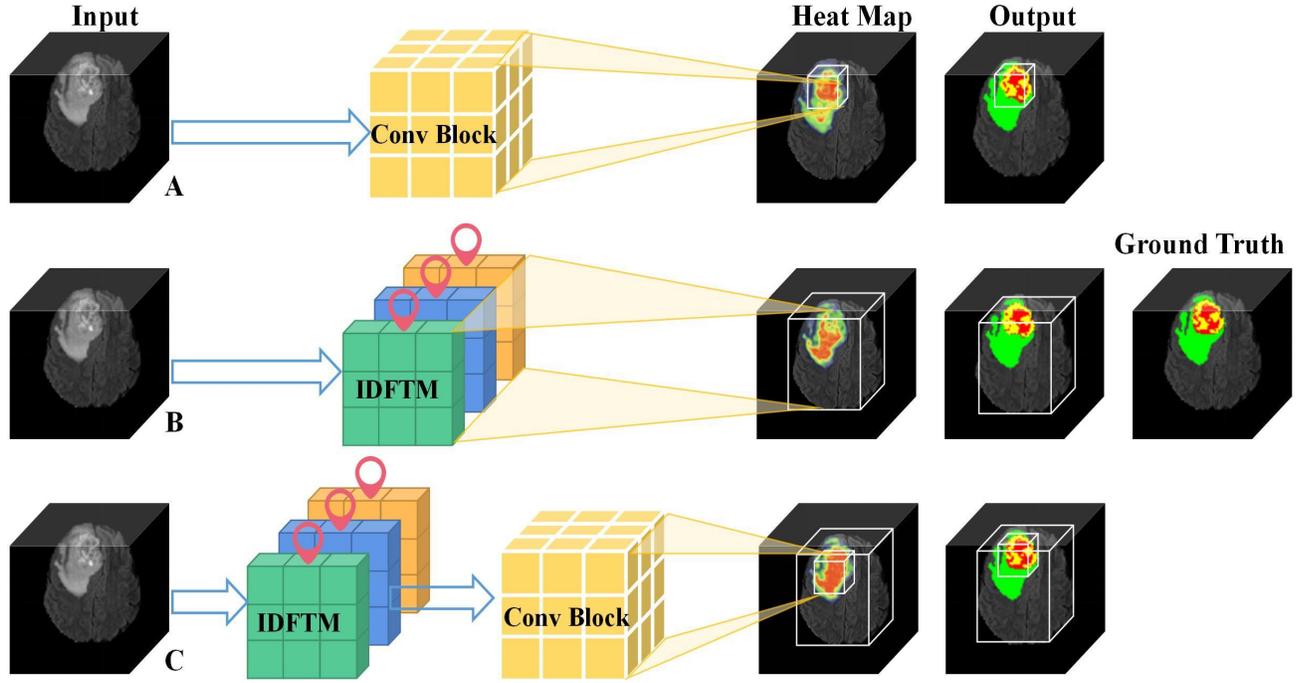

Fig. 1. The thermal maps of feature extraction by cascaded RBM, IDFTM and RBM+IDFTM respectively.

## II. METHODS

### A. Architecture Overview

In this paper, we make an attempt to develop 3D Brain tumor segmentation strategies with Transformer, combined with FHSA mechanism and IDFTM module, as is shown in Fig. 2. The overall framework of our proposed 3D Brainformer is composed of correlative steps, including the sequentialization, the encoder, and the decoder. Let there be totally $Z$ initial 3D MRI images of multiple modalities collected from scanners for subjects with brain tumors, $\{M_1, \cdots, M_z, \cdots, M_Z\}$, where $M_z (0 < z \leq Z)$ refers to the $z$ th 3D volume, of a uniform size $m_h \times m_w \times m_d$, on the axial, sagittal, and coronal planes. Each 3D MRI imaging $M_z$ is divided into $G$ blocks, $\{B_{z_1}, \cdots, B_{z_g}, \cdots, B_{z_G}\}$, of the size $H \times W \times D$ and $C$ channels, with $G = m_h \times m_w \times m_d / H \times W \times D$, where $B_{z_g}$ stands for the $g$ th block for the given 3D MRI imaging $M_z$. Suppose there are initially $\delta$ 3D MRI imaging blocks and the relevant labels as the pairs of the training samples for 3D Brainformer, with the training input set $B = \{B_1, \cdots, B_\lambda, \cdots, B_\delta\}$, $\lambda = 1, ..., \delta$. To be simplified, each 3D MRI imaging block $B_\lambda$ will be taken as one training input $X$. Specifically, we first try to establish 3D Brainformer architecture of $C$ channels, and feed 3D MRI imaging blocks of multiple modalities as the training inputs, with each 3D block of all modalities $X \in \mathbb{R}^{C \times H \times W \times D}$ at a spatial resolution of $H \times W \times D$. Such 3D MRI imaging blocks will be first reshaped into sequences during the sequentialization process, and then the baseline U-Net with Transformer will be inspired to construct the network. The sequences will be downsampled in the FHSA-based encoder to capture the improved 3D MRI encoding features at multiple scales, especially the spatial long-range dependencies among voxels, by the correlation transfer from layer to layer with the multi-head attention fusion of the proposed FHSA. Afterwards, the decoder will upsample the above 3D MRI contextual feature maps extracted from the encoder via skip connections, recover to the natively high spatial resolution, and finally output the segmentation feedback of higher accuracy, by utilizing IDFTM with cascaded 3D RBM. Here IDFTM develops a great breakthrough in original Transformer, with a plug-and-play attribute that could insert into any position of the network architecture, which allows for both capturing long-range dependencies and extracting fine-grained local feature maps, when cascading with 3D RBM, reducing the limitation of the sequentialized input of the fixed size and position during feature extraction in native Transformer.

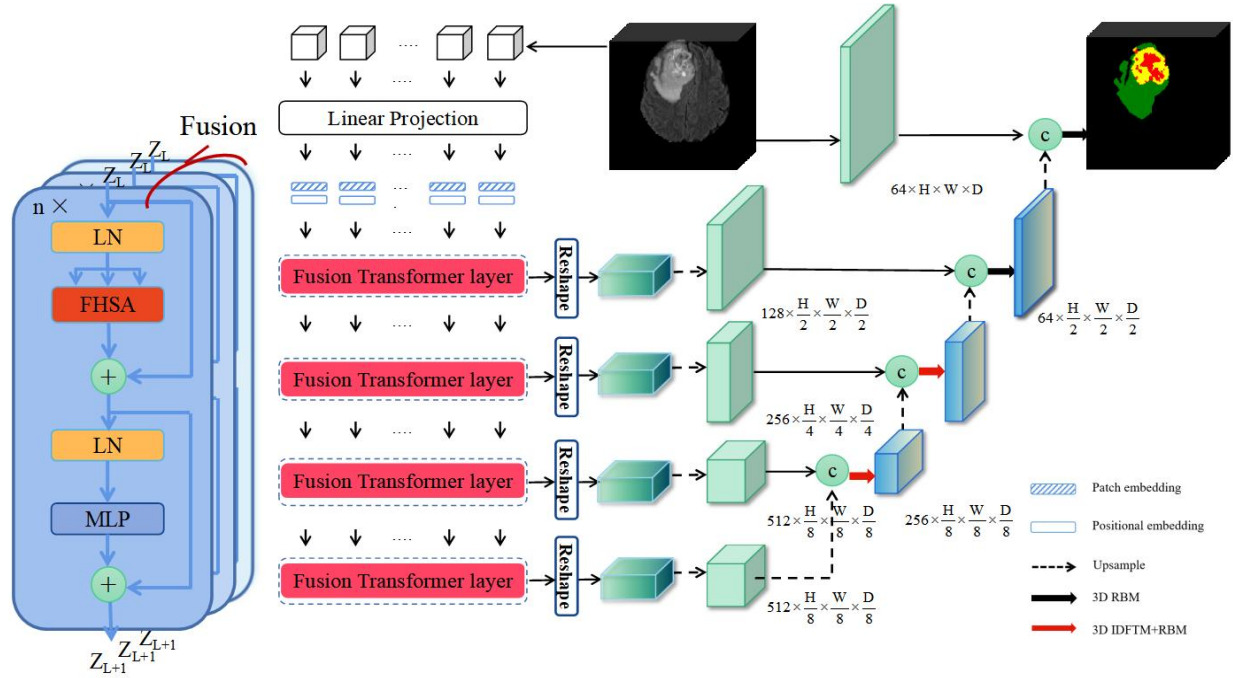

Fig. 2. The architecture of the proposed 3D Brainformer.

### B. Input Sequentialization

Each input $X$ will be detached into one non-overlapping 3D MRI sub-image block set $I = \{I_1,...,I_i,...,I_N\}$, where $N$ is the total number of sub-image blocks, and $I_i \in \mathbb{R}^{C \times p^3}$ represents the $i$ th 3D MRI sub-image block of the size $p \times p \times p$ with $C$ channels. One embedding layer will then transform each 3D MRI sub-image block $I_i$ into the $k$-dimensional space, with a linear projection function $F_X$:

$$S_\beta^i = F_X(I_i), i = 1,...,N \quad (1)$$

where $S_\beta^i$ refers to the $k$-dimensional embedding sequence, $S_\beta^i \in \mathbb{R}^k$.

The 1D positional embedding vector $S_e^i \in \mathbb{R}^k$ has been also added for each sequence to encode the spatial position, so the final sequence fed into the encoder is as follows:

$$S_0 = \{S_\beta^1 + S_e^1;...;S_\beta^N + S_e^N\} \quad (2)$$

where, $S_0 \in \mathbb{R}^{N \times k}$.

### C. Encoder and Decoder

In the encoder, we propose to establish Fusion Transformer layers to calculate self-attention metrics, which consists of FHSA and MLP modules, with Layernorm (LN) applied before each module, and residual connections after every module. More specifically, when the overall sequence $S_0$ derived from the sequentialization passes through Fusion Transformer layer by layer, the entire encoder with the total $L$ layers will be divided into concatenated sub-encoders at $n$ scales, each consisting of $L/n$ Fusion Transformer layers. All the sub-encoder output set $\{S_{(L \times b)/n} \in \mathbb{R}^{N \times k} \mid b = 1,...,n\}$ will be regarded as the multi-scale feature mapping of the subsequent prerequisite for the decoder.

In the decoder, the hidden layer sub-encoder output $S_{(L \times b)/n}$ will be reshaped into the subsequent prerequisite, i.e., the reconstructed 3D fused attention block set $T = \{T_1,...,T_b,...,T_n\}$ of the size $\mathbb{R}^{\frac{H}{p} \times \frac{W}{p} \times \frac{D}{p}}$ in the $k$-dimensional space. The above fused attention block will then be repeatedly up-sampled by multiple transposed convolution. For each up-sampled 3D attention block in $C_b$ channels, where $C_b = \rho / 2^{(n-b)}$, and $\rho$ denotes a certain number of the initial channels, mapping from the $k$-dimensional space, it will be concatenated at the higher dimension to produce a new 3D multi-scale fusion block $T'_b \in \mathbb{R}^{h \times w \times d}$, $h = H/p$, $w = W/p$, $d = D/p$. The cascaded use of one plug-and-play IDFTM and 3D RBM module will map such newly concatenated 3D multi-scale fusion block $T'_b$, to 3D fused feature map block $\hat{T}''_b$. The above decoding process is recursively repeated layer by layer, until the spatial resolution of 3D fused feature map block is restored to $H \times W \times D$, and finally output the segmentation result $Y$ of 3D MRI images for subjects with brain tumors.

### D. Fusion-Head Self-Attention Mechanism

Specifically, let the input sequence to FHSA module of the $l$ th layer be $S_h^l$. FHSA maps from $S_h^l$ to the corresponding output $S_h^{l+1}$ to capture multi-scale spatial context features.

Let there be $n_h$ attention heads in FHSA, the total self-attention mechanism from the input sequence $S_h^l$ could be derived from the following query vector set, i.e., the query

$Q_l$, the key $K_l$, the value $V_l$, composed of all the parallel attention heads, which could be determined by the following trainable weight mappings, $F_Q^l$, $F_K^l$, and $F_V^l$, respectively:

$$Q_l = \{Q_l^1, ..., Q_l^j, ..., Q_l^{n_h}\} = F_Q^l(S_h^l)$$
$$K_l = \{K_l^1, ..., K_l^j, ..., K_l^{n_h}\} = F_K^l(S_h^l) \quad (3)$$
$$V_l = \{V_l^1, ..., V_l^j, ..., V_l^{n_h}\} = F_V^l(S_h^l)$$

where, $Q_l^j, K_l^j, V_l^j \in \mathbb{R}^{N \times k_h}$, $j = 1, ..., n_h$, $k_h = k/n_h$ refers to the scaling factor of FHSA mechanism.

The attentional intensity $E_l^j$ will be computed for FHSA:

$$E_l^j = (Q_l^j (K_l^j)') / \sqrt{k_h} \quad (4)$$

where $E_l^j \in \mathbb{R}^{N \times N}$, and it divides the square root of the scale factor $k_h$ to prevent a too large inner product after matrix multiplication, with the transposition of $K_l^j$ as $(K_l^j)'$.

We propose to define one trainable attention logic mapping $F_A$ to execute the logical attention fusion from $n_h$ individual attention heads in parallel by the above attentional intensities, and obtain the fused head output $h_A$ with the logical association through the softmax layer:

$$h_A = Softmax(F_A([E_l^1, ..., E_l^j, ... E_l^{n_h}])) \quad (5)$$

where, $h_A \in \mathbb{R}^{n_h \times N \times N}$.

The trainable attention weight mapping $F_B$ is further put forward to fulfill the weight attention fusion from $n_h$ individual attention heads in parallel, and the fused head output $h_B$ with the weight association could be acquired:

$$h_B = F_B(h_A) \quad (6)$$

where, $h_B \in \mathbb{R}^{n_h \times N \times N}$.

The compound fused head output $h_v$ of self-attention could be formulated by the matrix multiplication with $\{V_l^1, ..., V_l^j, ..., V_l^{n_h}\} \in \mathbb{R}^{n_h \times N \times k_h}$:

$$h_v = h_B[V_l^1, ..., V_l^j, ..., V_l^{n_h}] \quad (7)$$

where, $h_v \in \mathbb{R}^{n_h \times N \times k_h}$.

The final output $S_h^{l+1}$ of FHSA will be computed by reprojecting the compound attention output $h_v$ back to $\mathbb{R}^{N \times k}$ space by the trainable weight mapping $F_P$:

$$S_h^{l+1} = F_P(h_v) \quad (8)$$

### E. Infinite Deformation Fusion Transformer Module

One primary challenge in Transformer is that its self-attention mechanism originates from the input sequentialization, which restricts to reasonably integrate with the conventional convolutions. We propose one IDFTM module to directly exert feature extraction on 3D multi-scale fusion blocks derived from encoder.

We first define three initially identical 3D convolution kernels, the query weight matrix $q_b^r$, the key weight matrix $k_b^r$, the value weight matrix $v_b^r$, to represent the attention weight matrices for the input $T_b'$ of IDFTM. The query set $Q_b$, the key set $K_b$, the value set $V_b$ will be attained to reconstruct the subsequent fusion self-attention respectively, by one sequentialization transformation function $F_u$:

$$Q_b = \{Q_b^1, ..., Q_b^r, ..., Q_b^{n_h'}\}, Q_b^r = F_u(q_b^r(T_b'))$$
$$K_b = \{K_b^1, ..., K_b^r, ..., K_b^{n_h'}\}, K_b^r = F_u(k_b^r(T_b')) \quad (9)$$
$$V_b = \{V_b^1, ..., V_b^r, ..., V_b^{n_h'}\}, V_b^r = F_u(v_b^r(T_b'))$$

where $Q_b^r, K_b^r, V_b^r \in \mathbb{R}^{c_b^r \times \lambda}$, $r = 1, ..., n_h'$, the number of attention head in IDFTM is $n_h'$.

Furthermore, three iterable position weights are denoted as $\alpha_w = \{\alpha_w^1, ..., \alpha_w^r, ..., \alpha_w^{n_h'}\}$, $\alpha_h = \{\alpha_h^1, ..., \alpha_h^r, ..., \alpha_h^{n_h'}\}$, $\alpha_d = \{\alpha_d^1, ..., \alpha_d^r, ..., \alpha_d^{n_h'}\}$, to learn the width-wise, the height-wise, and the depth-wise positional encoding for the newly concatenated 3D multi-scale fusion block $T_b'$, where $\alpha_w^r \in \mathbb{R}^{1 \times w \times 1 \times c_b^r}$, $\alpha_h^r \in \mathbb{R}^{h \times 1 \times 1 \times c_b^r}$, $\alpha_d^r \in \mathbb{R}^{1 \times 1 \times d \times c_b^r}$ represent the position weight parameters, $C_b^r = C_b / n_h'$ is the scale factor of IDFTM.

The fused spatial position weight matrix $\omega_\alpha^r$ for each corresponding attention head could be achieved:

$$\omega_\alpha^r = \alpha_w^r + \alpha_h^r + \alpha_d^r \quad (10)$$

where $\omega_\alpha^r \in \mathbb{R}^{h \times w \times d \times c_b^r}$, $\omega_\alpha = \{\omega_\alpha^1, ..., \omega_\alpha^r, ..., \omega_\alpha^{n_h'}\}$ consists of $n_h'$ fused spatial position weight matrices.

The above spatial position weight matrix $\omega_\alpha^r$ will be computed to reconstruct the sequence of spatial position weight $R_\omega^r$ to iteratively update the subsequent fusion self-attention mechanism, by the above sequentialization transformation function $F_u$:

$$R_\omega^r = F_u(\omega_\alpha^r) \quad (11)$$

where $R_\omega^r \in \mathbb{R}^{c_b^r \times \lambda}$, $\lambda = h \times w \times d$, $R_\omega = \{R_\omega^1, ..., R_\omega^r, ..., R_\omega^{n_h'}\}$ consists of $n_h'$ sequence of spatial position weights.

The attentional intensity $E_b^r$ will be attained for IDFTM:

$$E_b^r = (K_b^r)' Q_b^r + (R_\omega^r)' Q_b^r \quad (12)$$

where, $E_b^r \in \mathbb{R}^{\lambda \times \lambda}$, the transposition of $K_b^r$ is $(K_b^r)'$, and the transposition of $R_\omega^r$ is $(R_\omega^r)'$.

We define one trainable attention logic mapping $F_A'$ to perform the logical attention fusion from $n_h'$ individual attention heads in parallel by the attentional intensities, and

the fused head output $H_A$ with the logical association could be computed through the softmax layer:

$$H_A = Softmax(F'_A([E^1_b,...,E^r_b,...E^{n'_h}_b])) \quad (13)$$

where, $H_A \in \mathbb{R}^{n'_h \times \lambda \times \lambda}$.

The trainable attention weight mapping $F'_B$ is introduced to conduct the weight attention fusion from $n'_h$ individual attention heads in parallel, and the fused head output $H_B$ with the weight association could be received:

$$H_B = F'_B(H_A) \quad (14)$$

where, $H_B \in \mathbb{R}^{n'_h \times \lambda \times \lambda}$.

The compound fused head output $H_V$ of self-attention could be formulated by the matrix multiplication with $\{V^1_b,...,V^r_b,...,V^{n'_h}_b\} \in \mathbb{R}^{n'_h \times \lambda \times c^r_b}$:

$$H_V = H_B[V^1_b,...,V^r_b,...,V^{n'_h}_b] \quad (15)$$

where, $H_V \in \mathbb{R}^{n'_h \times \lambda \times c^r_b}$.

The final output $\hat{T}'_b$ of IDFTM will be calculated by reprojecting the compound attention output $H_V$ back to $\mathbb{R}^{C_b \times h \times w \times d}$ space by the trainable weight mapping $F_s$:

$$\hat{T}'_b = F_s(H_V) \quad (16)$$

So the total output with the cascaded 3D RBM module for the above feature extractor $\hat{T}'_b$ in the decoder will be:

$$\hat{T}''_b = \hat{T}'_b + \hat{R}(\hat{T}'_b) \quad (17)$$

where, $\hat{T}''_b \in \mathbb{R}^{C_b \times h \times w \times d}$, $\hat{R}$ denotes 3D RBM module.

## III. EXPERIMENTAL RESULTS

### A. Dataset

Our experiment was conducted with the publicly available benchmark of BraTS dataset. All the provided MRI images have undergone alignment, with $1.0 \times 1.0 \times 1.0$ mm³ isotropic resolution resampling and skull stripping. Specifically, BraTS 2019 consists of annotated training set, with 259 high grade gliomas (HGG) and 76 low grade gliomas (LGG) cases. BraTS 2018 is composed of annotated training set, with 210 HGG and 75 LGG cases, and one validation set of 66 subjects for online evaluation. BraTS 2017 includes annotated 285 subjects, with 46 subjects as the validation set for online evaluation. For each subject, the structural MRI imaging is made up of four modalities, i.e., the native T1-weighted (T1), the post-contrast T1-weighted (T1ce), the T2-weighted (T2) and the Fluid Attenuated Inversion Recovery (FLAIR). After pathological confirmation, MRI imaging of each subject has been annotated, with NCR and NET labeled as 1, ED labeled as 2, the values for ET labeled as 4, and the other regions labeled as 0. Further, the segmentation task is defined as three sub-regions, i.e., 1) TC including NCR, NET and ET, 2) ET sub region, 3) the whole tumor (WT) with the combination of TC and ED.

### B. Configuration Details

Our proposed model is programmed with PyTorch on a single NVIDIA RTX 2080Ti GPU. Adam optimizer is used to update the network weights with an initial learning rate of $10^{-3}$. Softmax Dice Loss is applied to train 3D Brainformer with L2 Norm for model regularization and a weight decay rate of $10^{-5}$. The uniform block size is set to $160 \times 160 \times 16$ voxels, with random cropping and a batch size of 2. The segmentation accuracy will be evaluated in terms of multiple metrics, including Dice, Sensitivity, Positive Predictive Values (PPV), Hausdorff95 (HD95) respectively.

### C. Ablation Studies

Ablation studies have been conducted by 6 variants of our proposed approach, which could be divided in accordance with the encoder and the decoder. The 2 variants of the encoder include: (a) basic Transformer with MHSA, (b) Transformer with the proposed FHSA. The 3 variants of the decoder involve: (c) the basic RBM+RBM, (d) IDFTM+IDFTM, (e) IDFTM+RBM.

In our ablation studies, all training samples are originated from Brats2018 training set, and the samples of 50 newly added subjects have been selected from Brats2019 training set with their annotations for our test set. Table 1 lists the quantitative comparison of 6 variants for 3D Brainformer, with the best segmentation performances in bold. It has been observed from Table 1 that FHSA performed much better than MHSA for brain tumor segmentation for all the evaluation metrics, provided the same configuration in the decoder, with RBM+RBM exerted. Similarly, FHSA outperformed MHSA for all the evaluation metrics, provided the cascades of only IDFTMs, and IDFTM+RBM in the decoder. As a result, the proposed FHSA mechanism addressed the challenges in establishing the cross-fusion of remote dependencies at multiple scales with the trainable attention logic mapping and attention weight mapping involved, which exceeds the original MHSA that only concerns on the independent calculation from each attention head in parallel.

It could be seen from Table 1 the configuration of IDFTM+RBM in the decoder achieved overally accurate segmentation performances, with the same self-attention mechanism exerted in the encoder. The reason why IDFTM+RBM received the sub-optimal performances during WT and ET segmentation process is most probable that IDFTM+IDFTM turned out to capture more large-scale mutual connections with higher number of voxels. For the TC region, IDFTM+RBM obtained better segmentation results for all the evaluation metrics than IDFTM+IDFTM. We believe that the residual convolution in RBM module played an essential role in capturing much detailed brain tumor features at local regions, which tends to be somehow less regarded in native Transformer. Therefore, our proposed model pursues the combination between IDFTM and RBM, although it might bring the sub-optimal evaluation for a wider range of brain tumor, while in most cases, the segmentation performances behaved much higher than the only cascades of IDFTM+IDFTM. On the other hand, the use of RBM+RBM fulfilled lower performances among all the evaluation metrics, compared to the other two counterparts.

TABLE I
PERFORMANCE COMPARISON OF 6 VARIANTS FOR 3D BRAINFORMER IN OUR ABLATION STUDIES.

| Encoder | Decoder | Dice | | | Sensitivity | | | PPV | | | HD95 | | |
|---|---|---|---|---|---|---|---|---|---|---|---|---|---|
| | | WT | TC | ET | WT | TC | ET | WT | TC | ET | WT | TC | ET |
| MHSA | RBM+RBM | 0.873 | 0.831 | 0.779 | 0.930 | 0.906 | 0.918 | 0.943 | 0.933 | 0.946 | 6.921 | 9.570 | 6.723 |
| | IDFTM+IDFTM | 0.878 | 0.844 | 0.784 | 0.937 | 0.910 | 0.926 | 0.948 | 0.938 | 0.957 | 6.499 | 9.399 | 6.095 |
| | IDFTM+RBM | 0.885 | 0.849 | 0.788 | 0.946 | 0.923 | 0.932 | 0.954 | 0.946 | 0.964 | 5.659 | 8.928 | 5.835 |
| FHSA | RBM+RBM | 0.887 | 0.844 | 0.786 | 0.941 | 0.923 | 0.928 | 0.952 | 0.943 | 0.957 | 6.542 | 9.252 | 6.139 |
| | IDFTM+IDFTM | **0.902** | 0.851 | 0.795 | 0.953 | 0.927 | **0.942** | **0.972** | 0.952 | 0.966 | 6.284 | 8.784 | 5.652 |
| | IDFTM+RBM | <u>0.892</u> | **0.859** | **0.806** | **0.954** | **0.938** | <u>0.939</u> | <u>0.965</u> | **0.955** | **0.974** | **4.602** | **7.663** | **4.799** |

The qualitative analysis of brain tumor segmentation for example MRI images with the combination of four modalities as inputs, is shown in Fig. 3, including the ground truth annotation, the segmentation results of 6 variants for 3D Brainformer in our ablation studies, and the original FLAIR slices for reference, with ED, ET, NCR/NET in green, yellow, red. In Fig. 3, the abbreviation of M, F, R, I represents MHSA, FHSA, RBM and IDFTM respectively. It has been indicated that the combination of FHSA with IDFTM+RBM achieved significant improvement in segmentation performances for brain tumors, not only compared to the combination of MHSA in original Transformer but also RBM+RBM in classical U-Net. Our proposed model has made a great progress to tackle the challenges in NCR and NET segmentation by replacing MHSA with FHSA in the encoder, as well as the use of IDFTM+RBM in the decoder, while for the large-scale regions, IDFTM+RBM received comparable segmentation performances to IDFTM+IDFTM.

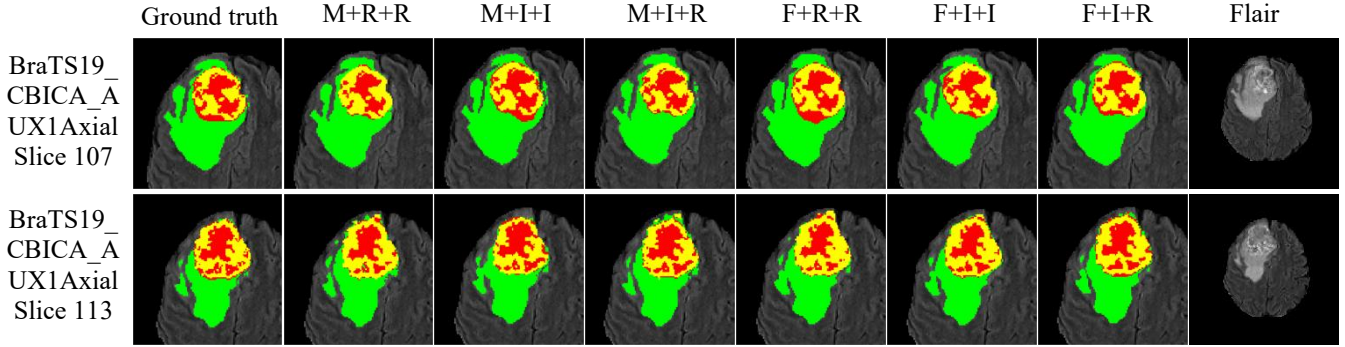

Fig. 3. Brain tumor segmentation results of 6 variants from 3D Brainformer for example MRI images with multiple modalities.

## D. Comparison With State-of-The-Art Approaches

Furthermore, we made a performance comparison of our proposed 3D Brainformer with the recently developed state-of-the-art approaches for brain tumor segmentation, including 3D U-Net, V-Net, and UNETR, as is listed in Table II, with the best segmentation performances in bold and the runner-up underlined. All the training samples are originated from Brats2018 and Brats2017 training sets, and the validation is from Brats2018 and Brats2017 validation sets. Note that the recently developed Medical Image Segmentation via Self-distilling TransUNet (MISSU) carried out their experiment on BraTS2019 training and validation datasets, which is different from our experimental setup. It has been demonstrated that our proposed model outperformed most of the state-of-the-art approaches on several evaluation metrics, and meanwhile achieved comparable segmentation results for the other evaluation metrics.

TABLE II
PERFORMANCE COMPARISON OF 3D BRAINFORMER WITH THE RECENTLY DEVELOPED STATE-OF-THE-ART APPROACHES

| Methods | Dice | | | Sensitivity | | | PPV | | | HD95 | | |
|---|---|---|---|---|---|---|---|---|---|---|---|---|
| | WT | TC | ET | WT | TC | ET | WT | TC | ET | WT | TC | ET |
| 3D U-Net [6] | 0.854 | 0.816 | 0.757 | 0.919 | 0.897 | 0.899 | 0.923 | 0.928 | 0.923 | 12.690 | 10.7285 | 9.832 |
| V-Net [7] | 0.863 | 0.822 | 0.756 | 0.922 | <u>0.909</u> | 0.905 | 0.926 | 0.931 | 0.932 | 9.646 | 11.769 | 8.725 |
| UNETR [10] | <u>0.873</u> | <u>0.831</u> | <u>0.779</u> | <u>0.930</u> | 0.906 | <u>0.918</u> | <u>0.943</u> | <u>0.933</u> | <u>0.946</u> | <u>6.921</u> | <u>9.570</u> | <u>6.723</u> |
| Ours(3D Brainformer) | **0.892** | **0.859** | **0.806** | **0.954** | **0.938** | **0.939** | **0.965** | **0.955** | **0.974** | **4.602** | **7.663** | **4.799** |

Fig. 4 displays 3D Brain tumor segmentation results for example MRI images with the combination of modalities, including the ground-truth, the segmentation results of 3D U-Net, V-Net and our model, as well as the original FLAIR slices for reference, from left to right, respectively, with ED, ET, NCR/NET in green, yellow, red.

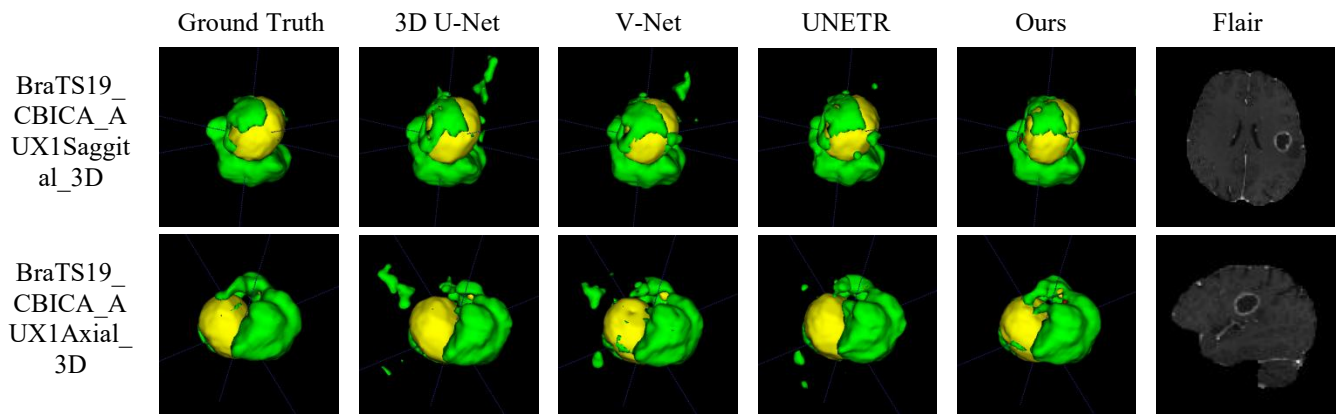

Fig. 4. Brain tumor segmentation results of 3D Brainformer and the state-of-the-art methods for example MRI images with multiple modalities.

It has been illustrated from Fig. 4 that our proposed model could elaborate in delicately localizing the target brain tumor in high precision, leading to less uncertainty, which is of great importance to the overall segmentation performances, compared to 3D U-Net and V-Net. Our proposed model successfully exploited the spatially long-range dependencies and inter-layer structures of brain tumors. FHSA mechanism paid more attention to calculate the multi-scale cross-fusion of the previously independent attention heads, provided the compound projection of self-attention matrix in the encoder. The significant improvement in mis-segmentation for ED regions has been detected, compared to the other approaches, showing its efficiency and robustness in depicting the long-range spatial dependencies.

Our developed scheme with IDFTM employed in the decoder also displays the remarkable improvements in segmentation performances, for NCR and NET regions. It has been indicated that our model not only concerned on the large spatial context of brain tumors, but also surpassed the convolution-based models only with the local feature extraction, via the cascaded IDFTM+RBM, which meanwhile emphasized on the detailed feature mapping for brain tumors. This is due to the fact that IDFTM is not restricted by the input sequentialization for self-attention mechanism in original Transformer, and could be flexibly inserted in any position of the network architecture, in a plug-and-play manner. While the original Transformer focuses more on the buildup of the long-range relationships in spatial contexts, and somehow underestimates the local detailed feature extraction from convolution. On the other hand, it is so difficult in accurately capturing the multi-scale compound self-attention if only convolution-based models are applied.

We further quantitatively assessed the overall segmentation performance of our proposed 3D Brainformer, from the mean segmentation results validated by online evaluation server in BraTS2017 and BraTS2018 challenges, as is reported in Table III, Table IV, on both slice-wise and volume-wise, with the best segmentation performances in bold, the runner-up underlined, the final winner marked '*', unreported results marked '-'. It has been demonstrated from Table III that our proposed model won ET and TC region with the mean Dice higher than the reported champion in BraTS2017 on volume-wise respectively. In term of the mean HD95, our proposed model outperformed all the above slice-wise methods for ET and WT regions, and achieved a sub-optimal segmentation result for TC regions, compared to the champion of the volume-wise methods. It has been shown from Table IV that our proposed model received better performances for WT, TC, ET from mean Dice, compared to the highest score of the slice-wise methods in BraTS2018, i.e., Spatial-ConvNet [13] and MCCNN [14]. In comparison of volume-wise methods, 3D Brainformer achieved a slightly lower sub-optimal segmentation result than the reported champion in BraTS2018 for ET, TC region on Dice, and ET region on HD95. Note that the auto-encoder regularization proposed by Myronenko [15] received the best segmentation performance in most of the evaluation metrics, while it set up an additional branch from the auto-encoder to regularize the backbone and reconstruct the input 3D MRI imaging. This auto-encoder branch greatly enhanced the feature extraction of the encoder backbone. In our 3D Brainformer framework, we only made use of L2 regularization to learn the network parameters without any additional branches involved, which further mitigates the computational costs. The time period for each subject in 3D Brainformer is roughly 2-4 minutes, which could be well suited for the upcoming clinical applications. Generally speaking, our proposed model has been proved to be quite competitive in segmentation performances with high accuracy, in terms of several evaluation metrics, compared to the state-of-the-art methods in both BraTS2017 and BraTS2018 challenges.

TABLE III
PERFORMANCE COMPARISON OF 3D BRAINFORMER WITH THE STATE-OF-THE-ART METHODS ON BRATS 2017 VALIDATION DATASET.

| Types | Methods | Dice | | | HD95 | | |
|---|---|---|---|---|---|---|---|
| | | ET | WT | TC | ET | WT | TC |
| Slice-wise | Roy et al. [16] | 0.716 | 0.892 | 0.793 | 6.612 | 6.735 | 9.806 |
| | Islam et al. [17] | 0.689 | 0.876 | 0.761 | 12.938 | 9.820 | 12.361 |
| | Jungo et al. [18] | 0.749 | **0.901** | 0.790 | **5.379** | **5.409** | **7.487** |
| | Lopez et al. [18] | 0.567 | 0.783 | 0.685 | 23.828 | 30.316 | 38.077 |
| | Shaikh et al. [20] | 0.650 | 0.870 | 0.680 | - | - | - |
| | Zhao et al. [21] | **0.754** | 0.887 | **0.794** | - | - | - |
| Volume-wise | Isensee et al. [22] | 0.732 | 0.896 | 0.797 | 4.550 | 6.970 | 9.480 |
| | Jesson et al. [23] | 0.713 | 0.899 | 0.751 | 6.980 | **4.160** | 8.650 |
| | Kamnitsas et al. [24]* | 0.738 | 0.901 | 0.797 | 4.500 | 4.230 | **6.560** |
| | Li et al. [25] | 0.704 | 0.871 | 0.682 | 7.699 | 10.396 | 13.062 |
| | Pereira er al. [26] | 0.719 | 0.889 | 0.758 | 5.738 | 6.581 | 11.100 |
| | Castillo et al. [27] | 0.690 | 0.860 | 0.690 | - | - | - |
| | Ours | **0.742** | 0.894 | **0.803** | 4.687 | 4.793 | 7.826 |

TABLE IV
PERFORMANCE COMPARISON OF 3D BRAINFORMER WITH THE STATE-OF-THE-ART METHODS ON BRATS 2018 VALIDATION DATASET.

| Types | Methods | Dice | | | HD95 | | |
|---|---|---|---|---|---|---|---|
| | | ET | WT | TC | ET | WT | TC |
| Slice-wise | Hu et al. [14] | 0.718 | **0.882** | 0.748 | 5.686 | 12.607 | 9.622 |
| | Chandra et al. [28] | 0.741 | 0.872 | 0.799 | 5.575 | 5.038 | 9.588 |
| | Banerjee et al. [13] | **0.772** | 0.880 | **0.801** | **4.290** | **4.900** | **6.590** |
| | Carver et.al [29] | 0.710 | 0.880 | 0.770 | 4.460 | 7.090 | 9.570 |
| | Salehi et al. [30] | 0.704 | 0.822 | 0.733 | 9.668 | 9.610 | 13.909 |
| | Chen et al. [31] | 0.707 | 0.845 | 0.731 | 10.385 | 11.822 | 15.066 |
| Volume-wise | Nuechterlein et al. [32] | 0.737 | 0.883 | 0.814 | 5.295 | 5.461 | 7.850 |
| | Cabezas et al. [33] | 0.740 | 0.889 | 0.726 | 5.304 | 6.956 | 11.924 |
| | Gates et al. [34] | 0.678 | 0.806 | 0.685 | 14.523 | 14.415 | 20.017 |
| | Puch et al. [35] | 0.758 | 0.895 | 0.774 | 4.502 | 10.656 | **7.103** |
| | Weninger et al. [36] | 0.712 | 0.889 | 0.758 | 8.628 | 6.970 | 10.910 |
| | Myronenko [15] * | **0.816** | **0.904** | **0.860** | 3.805 | 4.483 | 8.279 |
| | Chen et al. [37] | 0.749 | 0.894 | 0.831 | 4.432 | 4.716 | 7.748 |
| | Ours(3D Brainformer) | 0.778 | 0.897 | 0.823 | 4.364 | 4.883 | 8.575 |

## IV. DISCUSSION AND CONCLUSION

In this paper, we have put forward a novel deep learning based framework for the brain tumor segmentation tasks in MRI imaging, called 3D Brainformer. Our developed scheme have addressed challenges in classic convolution-based approaches. First, the nature of the convolutional architectures tends to inherently focus on the local receptive fields among adjacent voxels, which makes it confined to the semantically global spatial contexts in segmentation for MRI imaging. Specifically, our proposed model employs Transformer in the context of conventional U-Net architecture instead, to leverage the limitations of capacities in convolutional networks with a lack of explicitly encoding long-range dependencies. Since the majority of current Transformer-based methods are performed with 2D MRI slices, we upgrade our segmentation model by altering into the architecture for 3D volumetric data. Second, although the original Transformer allows us to establish global long-range connections, it is still difficult to completely incorporate the self-attention mechanism from each independent attention head at multiple scales among layers in MHSA. So we present a novel FHSA mechanism to enable the compound cross-fusion feature extraction through attention logic and weight mapping, and enhance the inter-layer self-attention at multiple scales for MRI imaging. Third, given the limitations of input sequentialization in Transformer, we innovatively create IDFTM module to seamlessly integrate into any network architecture, which extremely overcomes the challenge of combining Transformer with convolution-based networks in a plug-and-play manner. In this way, we construct the cascaded IDFTM with RBM to globally capture the spatial context at multi-scales with long-range dependencies and meanwhile accurately extract the details from local receptive fields of brain tumors. It has been demonstrated from our experiment results that our proposed scheme has outperformed most of the state-of-art approaches for brain tumor segmentation on the public BRATS datasets, in terms of Dice, sensitivity, PPV, HD95 metrics. We make a preliminary study on the potential 3D Transformer design characterized by the infinite deformable fusion modules in this work, which permits us to delicately distinguish the functionality of both large-scale self-attention mechanism and the feature mapping from local receptive field. This study demonstrates the premise in quantifying brain tumor segmentation categories with deep learning framework and exhibiting spatially fine-grained segmentation patterns in clinical applications.